\documentclass[]{article}
\usepackage{amsmath}
\usepackage{amssymb}
\usepackage{graphics}
\usepackage{graphicx}
\usepackage{epsfig}
\usepackage{dcolumn}
\usepackage{bm}
\usepackage{lineno}
\usepackage{multirow}
\usepackage{ragged2e}

\title{\bf Study of charging-up effect for a single mask triple GEM detector}
\date{}
\begin{document}
		\maketitle
		\vspace*{-1cm}
		\centering
		{
		\author{S.~Chatterjee\footnote[1]{Corresponding Author:
sayakchatterjee896@gmail.com, sayakchatterjee@jcbose.ac.in},}
		\author{A.~Sen,}
		\author{S.~Das,}
		\author{S.~K.~Ghosh,}
		\author{S.~Biswas}
		}
		\vspace*{0.5cm}
		
		{Department of Physics and Centre for Astroparticle Physics and Space Science (CAPSS), Bose Institute, EN-80, Sector V, Kolkata-700091, India}
		
		\vspace*{0.5cm}
		\centering{\bf Abstract}
		\justify
			With the advancement of the accelerator systems and the requirements of high luminosity particle beams to reach different physics goals, detectors with good position resolution and high rate handling capability have become essential for designing any High Energy Physics~(HEP) experiments. The Gas Electron Multiplier~(GEM) detectors are widely used in many HEP experiments as a tracking device because of their good spatial resolution and rate handling capability. 
						
			The presence of the dielectric medium inside the active volume of the GEM detector changes its behaviour when exposed to external radiation. This mechanism is commonly referred as the charging-up effect. In this article, the effect of the charging-up phenomenon and the initial polarisation effect of the dielectric on the gain of the chamber are reported for a single mask triple GEM chamber with Ar/CO$_2$ gas mixture. 
			
		\vspace*{0.25cm}
		Keywords: Gas Electron Multiplier~(GEM); Single mask foil; High Voltage~(HV);  Charging-up; Polarisation; Gain

	\section{Introduction}\label{intro}
	
	Gas Electron Multiplier~(GEM) detector, introduced by Fabio Sauli in 1997~\cite{sauli_GEM}, is used in many HEP experiments~\cite{gem_review} for its high rate handling capability ($\sim$1MHz/mm$^2$)~\cite{sauli_GEM} and good position resolution ($\sim$ 70 $\mu$m)~\cite{ketzer}. The micro pattern structure of the foil is utilised to achieve good position resolution, to reduce the ion backflow in case it is used in a drift chamber and also to handle high rates~\cite{CMS_upgrade,alice_upgrade,cbm_detector_system,s_biswas_spark,s_chatterjee_spark}. A standard GEM foil consists of a 50~$\mu$m Kapton foil with 5~$\mu$m copper cladding on both sides. A large number of holes are etched into the Copper cladded Kapton foil using the photolithographic technique~\cite{GEM_foil}. Depending on the etching technique, the GEM foils are classified as a double mask or single mask type. The single mask technique is developed mainly due to the requirement of the large area GEM chambers where the alignment of the two masks (on the top and bottom side of the copper cladded Kapton foil) is not possible. However, the holes obtained with the single mask technique is asymmetrically bi-conical in shape as compared to the holes obtained with the double mask technique~\cite{pinto_large_area_gem}. Different studies are performed to understand the effect of the variation of hole geometry on the final performance of the chamber~\cite{gem_hole_geometry_asym_1,s_das,gem_hole_geometry_asym_2}. The dielectric medium~(Kapton) present in the active volume of the detector, changes the behaviour of the chamber when exposed to external radiation. As a result, the gain of the chamber increases initially and then reaches a constant value asymptotically. This increase in gain is due to the charging-up of the dielectric medium. Many different groups have reported on the studies to understand the effect of this charging-up phenomenon in GEM detectors~\cite{charging_up_philip,charging_up_azmoun,charging_up_alfonsi}. The charging-up effect for a double mask triple GEM prototype~(10~cm$~\times$~10~cm) is reported in Ref.~\cite{s_chatterjee_charging_up}. In the present work, the effect of charging-up on a single mask triple GEM chamber of dimension 10~cm~$\times$~10~cm, operated with Ar/CO$_2$ gas mixture in 70/30 volume ratio is investigated with Fe$^{55}$ X-ray source and reported in this article. The effect of initial polarisation of the dielectric is also investigated for different gain of the chamber with different irradiation rates. The details of the GEM prototype and experimental setup are discussed in section~\ref{set_up} and the results are reported in section~\ref{res}.   
	
	\section{Detector description and experimental setup}\label{set_up}
	The single mask triple GEM detector prototype consisting of 10~cm~$\times$10~cm standard stretched foils, obtained from CERN is assembled in the clean room of the RD51 laboratory~\cite{RD51}. 
	%%%%%%%%%%%%%%%%%%%%%%%%%%%%%%%%%%%%%%%%%%%%%%%%%%%
	\begin{figure}[htb!]
		\begin{center}
			\includegraphics[scale=0.30]{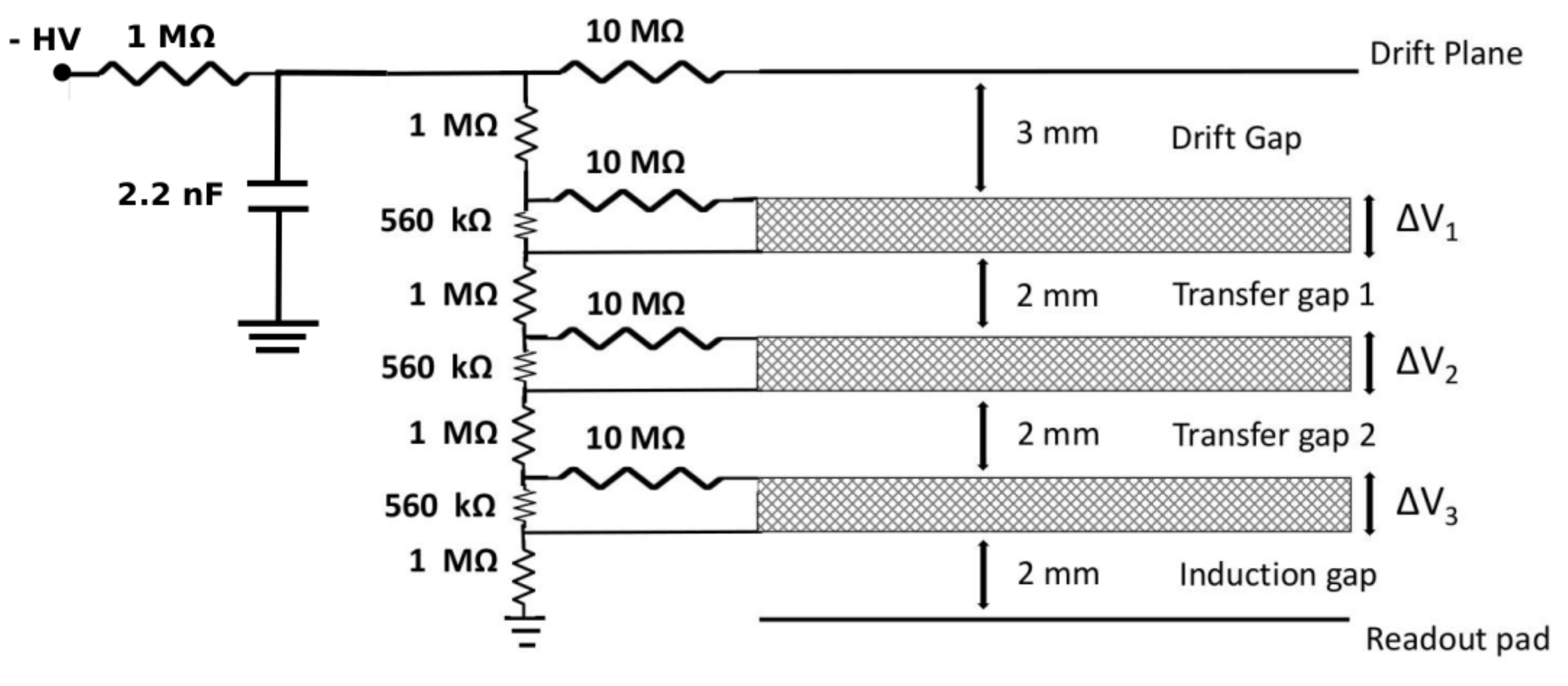}
			\caption{\label{setup} Schematic of the HV distribution through the resistive chain to different planes of the single mask triple GEM detector. A low pass HV filter is used between the HV line and resistive chain. }\label{fig1}
		\end{center}
	\end{figure}
	%%%%%%%%%%%%%%%%%%%%%%%%%%%%%%%%%%%%%%%%%%%%%%%%%%%
	The drift gap, transfer gap~1, transfer gap~2 and the induction gap of the chamber is kept at 3, 2, 2, 2~mm respectively~\cite{rama_adak}.  A voltage divider resistive chain is used to power the chamber as shown in Fig.~\ref{fig1}. A low pass filter is placed between the HV module and resistor chain as shown in Fig.~\ref{fig1} to bypass the ac components present in the HV line. The chamber has an XY printed board (256 X-tracks, 256 Y-tracks) on the base plate and that works as the readout plane. Each of 256 X-tracks and 256 Y-tracks is connected to two 128 pin connectors. However, for the purpose of this work individual track readout is not used. Instead of that, a sum-up board (provided by CERN) is used for each 128 pin connector. A total of 4 sum-up boards are used in this prototype. The signal from one of the sum-up boards is put to a charge sensitive preamplifier (VV50-2) having gain~2~mV/fC and shaping time 300~ns~\cite{preamplifier}. The output signal from the preamplifier is fed to a linear Fan-in-Fan-out (linear FIFO) module. One analog signal from the linear FIFO is put to a Single Channel Analyser (SCA) to measure the rate of the incident particle. The SCA is operated in integral mode and the lower level in the SCA is used as the threshold to the signal. The threshold is set at 0.9~V to reject the noise. The discriminated signal from the SCA, which is TTL in nature, is put to a TTL-NIM adapter and the output NIM signal is counted using a scaler. Another output of the linear FIFO is fed to a Multi-Channel Analyser (MCA) to obtain the energy spectra. The schematic of the electronic circuit is shown in Fig.~\ref{fig2}.
	%%%%%%%%%%%%%%%%%%%%%%%%%%%%%%%%%%%%%%%%%%%%%%%%%%%
	\begin{figure}[htb!]
		\begin{center}
			\includegraphics[scale=0.40]{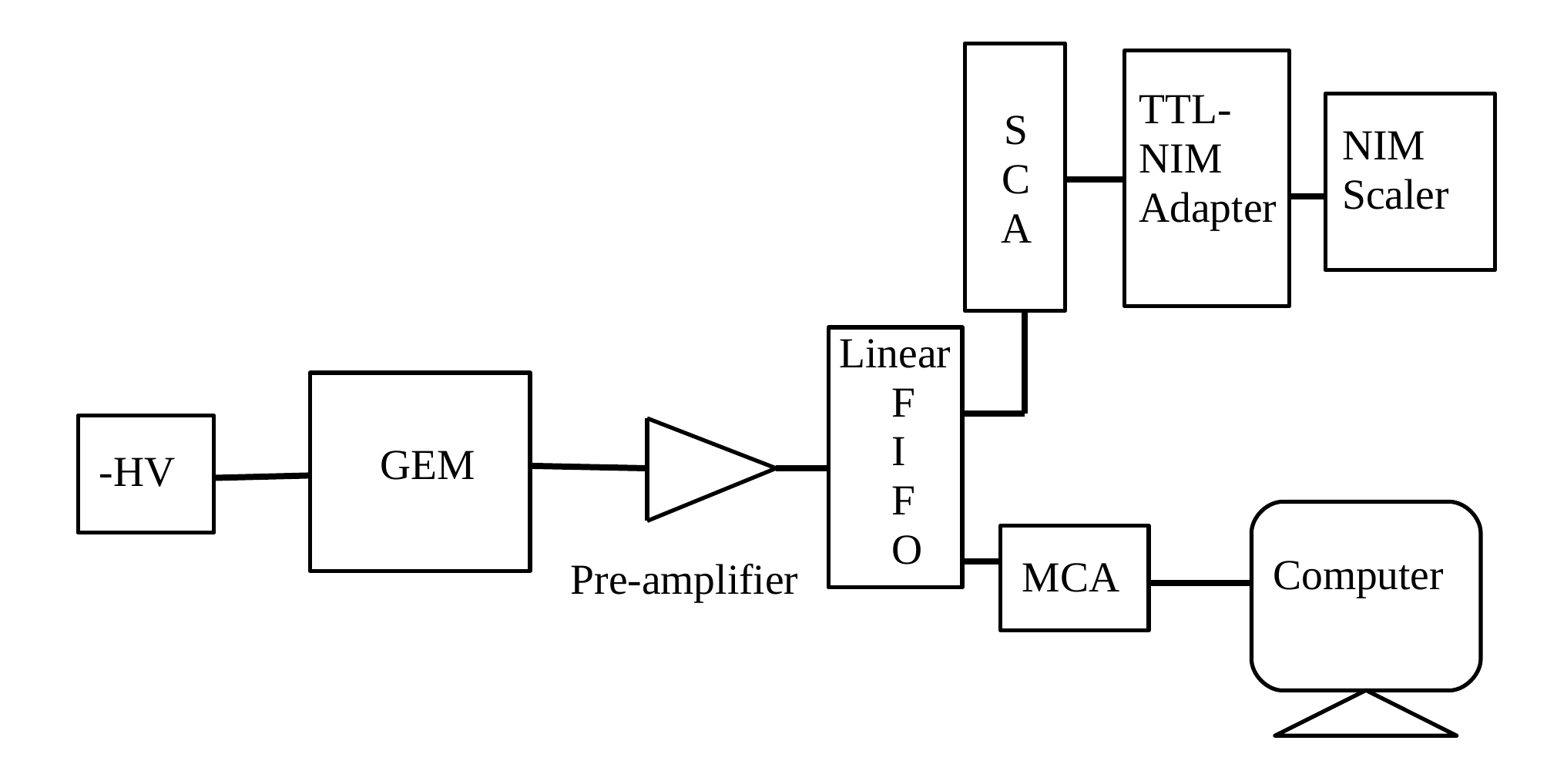}
			\caption{Schematic of the electronic circuit used for data acquisition.}\label{fig2}
		\end{center}
	\end{figure}
	%%%%%%%%%%%%%%%%%%%%%%%%%%%%%%%%%%%%%%%%%%%%%%%%%%%
	For the entire study, the chamber is operated with pre-mixed Ar/CO$_2$ gas in a 70/30 volume ratio. A constant gas flow rate of $\sim$3.5 l/hr is maintained using a V\"ogtlin gas flow meter. Collimators are used to irradiate the chamber with different X-ray flux coming from the Fe$^{55}$ source. The ambient temperature, pressure and relative humidity are monitored continuously using a data logger, built-in house~\cite{data_logger}.
	
	\section{Results}\label{res}
\begin{figure}[htb!]
		\begin{center}
			\includegraphics[scale=0.40]{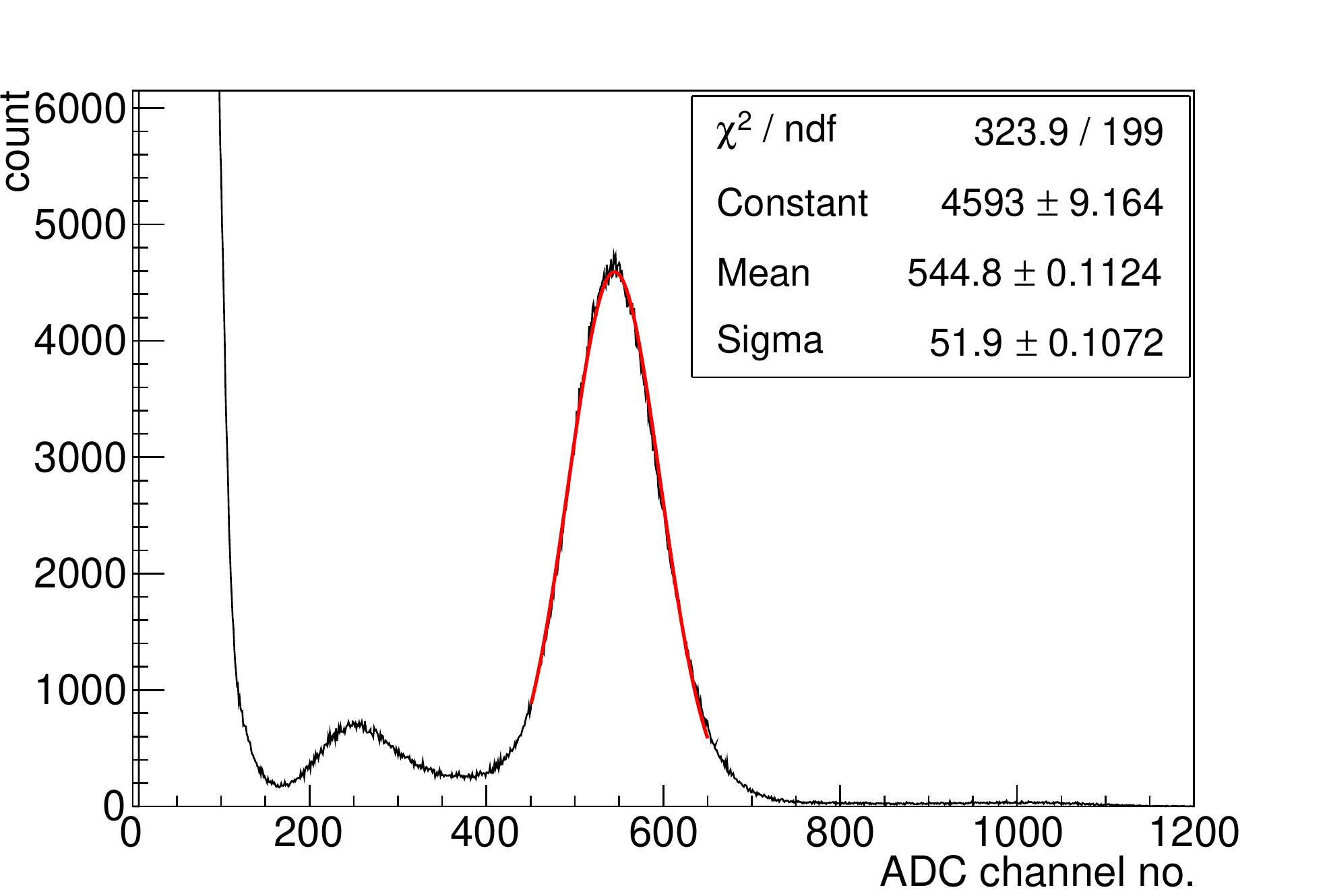}
			\caption{Typical Fe$^{55}$ spectra at - 5100 V. The $\Delta V$ across each of the GEM foil is 410 V. The corresponding drift field, transfer fields and induction field are $\sim$~2.4 kV/cm, $\sim$ 3.7 kV/cm and $\sim$~3.7~kV/cm respectively.}\label{fig3}
		\end{center}
	\end{figure}

	The effect of charging-up of the dielectric inside the active volume of the chamber on its performance is studied. The Fe$^{55}$ energy spectra obtained from the MCA is analysed and the gain is calculated in the same way as mentioned in Ref.~\cite{s_chatterjee_charging_up}. In Fig.~\ref{fig3}, the typical Fe$^{55}$ spectrum is shown for the single mask triple GEM chamber at - 5100 V which corresponds to $\Delta V$ of 410 V across  
	%%%%%%%%%%%%%%%%%%%%%%%%%%%%%%%%%%%%%%%%%%%%%%%%%%%
	%%%%%%%%%%%%%%%%%%%%%%%%%%%%%%%%%%%%%%%%%%%%%%%%%%%
	each of the GEM foils. To measure the initial polarisation effect of the dielectric, the recording of the spectra is started as soon as the HV reached its specific set value and the source is placed at a particular position of the chamber. The same Fe$^{55}$ source is used to irradiate the chamber as well as to record the X-ray spectra. To see the effect of HV and irradiation rates on the initial polarisation of the dielectric, X-ray spectra with different HV and collimator settings are recorded for 20~seconds without any interval between two consecutive measurements. The details of the HV settings used for the study of initial polarisation effect, their corresponding $\Delta V$ across each GEM foils, average gain and the electric field strengths in the drift, transfer and induction gaps are listed in Table~\ref{table1}. 
	
\begin{figure}[htb!]
		\begin{center}
			\includegraphics[scale=0.40]{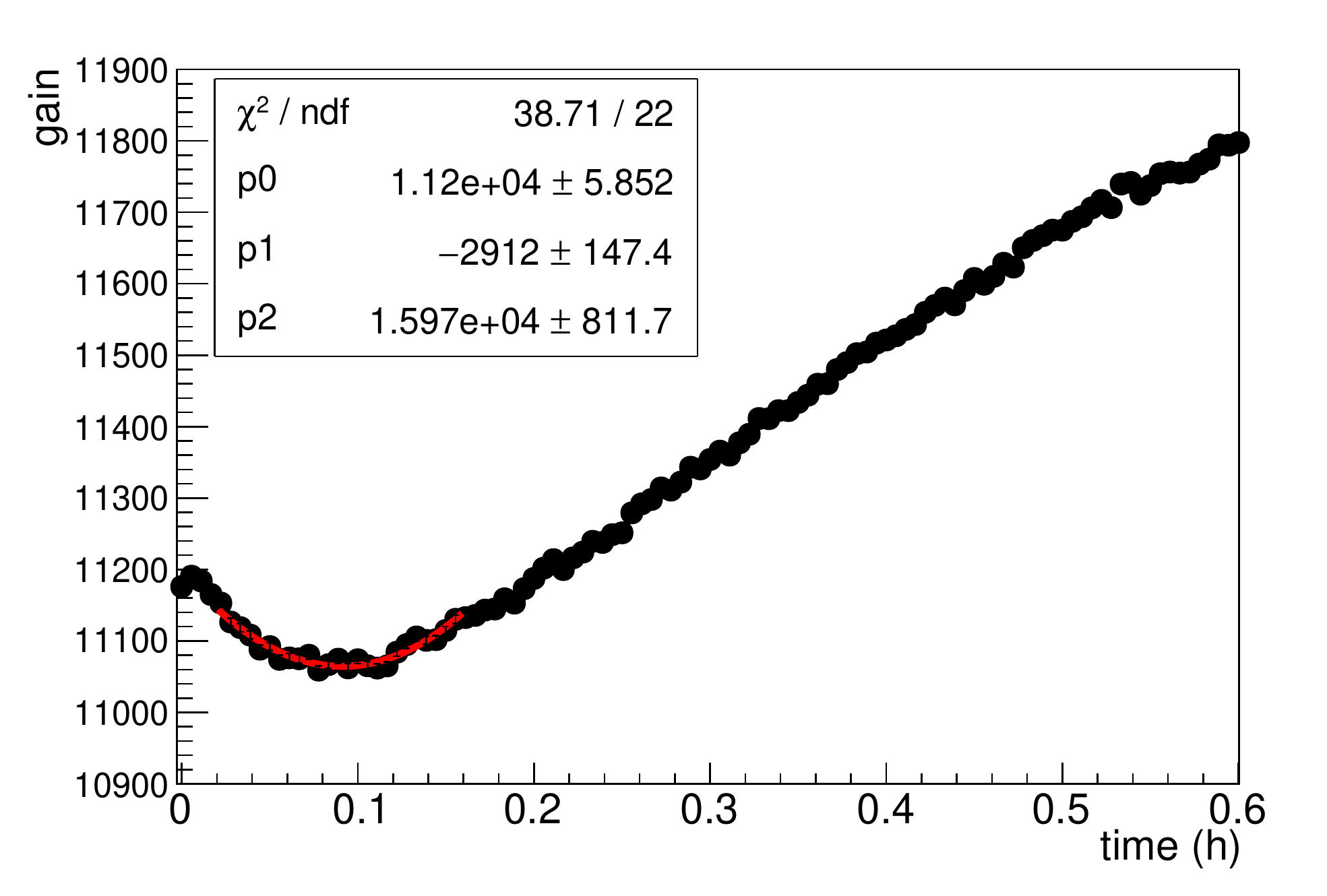}
			\caption{Variation of the gain as a function of time at a HV of - 5100 V. The initial decrease in gain due to the polarisation effect is fitted with a 2$^{nd}$ degree polynomial. }\label{fig4}
		\end{center}
	\end{figure}

	%%%%%%%%%%%%%%%%%%%%%%%%%%%%%%%%%%%%%%%%%%%%%%%%%%%%%%%%%%%%%%%%%%%
	\begin{center}
		
		\begin{table}[htb!]
			\begin{center}
				
				\vspace*{0.4cm}
				\resizebox{\columnwidth}{!}{
					\begin{tabular}{|c|c|c|c|c|c|} \hline
						HV & $\Delta V$ & Gain & Drift field & Transfer field & Induction field \\ 
						(V)&(V)&  &(kV/cm)&(kV/cm)&(kV/cm)\\ \hline
						- 5085 & 409 &$\sim$ 12950 &$\sim$ 2.4  & $\sim$ 3.6& $\sim$ 3.6 \\ \hline
						- 5100 & 410 &$\sim$ 13600 & $\sim$ 2.4  & $\sim$ 3.7& $\sim$ 3.7 \\ \hline
						- 5115 &  411 &$\sim$ 14300 &$\sim$ 2.4  & $\sim$ 3.7& $\sim$ 3.7\\ \hline
					\end{tabular}
				}
				\caption{Potential difference across each GEM foil, average gain and fields on the various gaps of the triple GEM chamber for different HV settings. The gain values are measured with an irradiation rate of $\sim$~0.14 kHz/mm$^2$.} \label{table1}
			\end{center}
		\end{table}
		
	\end{center}
	%%%%%%%%%%%%%%%%%%%%%%%%%%%%%%%%%%%%%%%%%%%%%%%%%%%%%%%%%%%%%%%%%%%
	Due to the initial polarisation effect, the decrease in gain for the first few minutes as reported in Ref.~\cite{s_chatterjee_charging_up}, is also observed in this study. In Fig.~\ref{fig4}, the variation of the gain as a function of time is shown for a particle flux of $\sim$ 0.14 kHz/mm$^2$ at a HV of - 5100~V. The variation in ambient temperature~(T) to pressure~(p) ratio for the first 20 minutes is below 1$\%$ for all the measurements. Though the variation of gain in any gaseous detector with temperature and pressure is a well known phenomenon~\cite{tp_gem} but since the variation in temperature to pressure ratio is small, no T/p normalisation is performed for this initial period. To identify the time up to which the gain decreases initially, the gain is fitted with a 2$^{nd}$ degree polynomial using the chi-square minimisation technique, as available in ROOT~\cite{cern_root}. From the fitting parameters, the ratio \textit{p1}/2\textit{p2} gives the minimum. The fitted curve with the respective chi-square value is shown in Fig.~\ref{fig4}. 
	%%%%%%%%%%%%%%%%%%%%%%%%%%%%%%%%%%%%%%%%%%%%%%%%%%%
	%%%%%%%%%%%%%%%%%%%%%%%%%%%%%%%%%%%%%%%%%%%%%%%%%%%
	The same technique is repeated for different gain and rate configurations to find out the time up to which the gain reduces due to the initial polarisation of the dielectric and then again starts to increase due to the charging-up effect.   
	
	%%%%%%%%%%%%%%%%%%%%%%%%%%%%%%%%%%%%%%%%%%%%%%%%%%%%%%%%%%%%%%%%%%%
	\begin{table}[htb!]
		\centering
		\resizebox{\columnwidth}{!}{
			\begin{tabular}{|c|c|c|c|c|}
				\hline
				\multirow{2}{*}{$\Delta$V } & \multicolumn{4}{c|}{time (h)} \\ \cline{2-5} 
				& rate $\sim$~0.04  & rate $\sim$~0.14 & rate $\sim$~0.42 & rate $\sim$~7.78  \\ 
				(V)& (kHz/mm$^2$)& (kHz/mm$^2$)& (kHz/mm$^2$)& (kHz/mm$^2$) \\ \hline
				409	& 0.133 & 0.110  & 0.115     &0.126      \\ 
				&(\underline{+} 0.018)    &(\underline{+} 0.013)   &(\underline{+} 0.005)&(\underline{+} 0.004) \\ \hline
				
				410& 0.115     & 0.098    & 0.103     &0.097    \\ 
				&(\underline{+} 0.018)&(\underline{+} 0.006) &(\underline{+} 0.003)& (\underline{+} 0.004)\\ \hline
				
				411& 0.128    & 0.091  & 0.075    &0.076  \\ 
				&(\underline{+} 0.019) &(\underline{+} 0.010)   &(\underline{+} 0.004) &(\underline{+} 0.003)\\ \hline
			\end{tabular}
		}
		\caption{ Variation of time in hour up to which the initial gain decreases with different $\Delta V$ and irradiation rates.} \label{table2}
		
	\end{table}
	%%%%%%%%%%%%%%%%%%%%%%%%%%%%%%%%%%%%%%%%%%%%%%%%%%%%%%%%%%%%%%%%%%%
	In table~\ref{table2}, the time~(hour) up to which the initial gain decreases due to the initial polarisation effect is listed for different $\Delta V$ and irradiation rates. 
	
	%%%%%%%%%%%%%%%%%%%%%%%%%%%%%%%%%%%%%%%%%%%%%%%%%%%
	\begin{figure}[htb!]
		\begin{center}
			\includegraphics[scale=0.55]{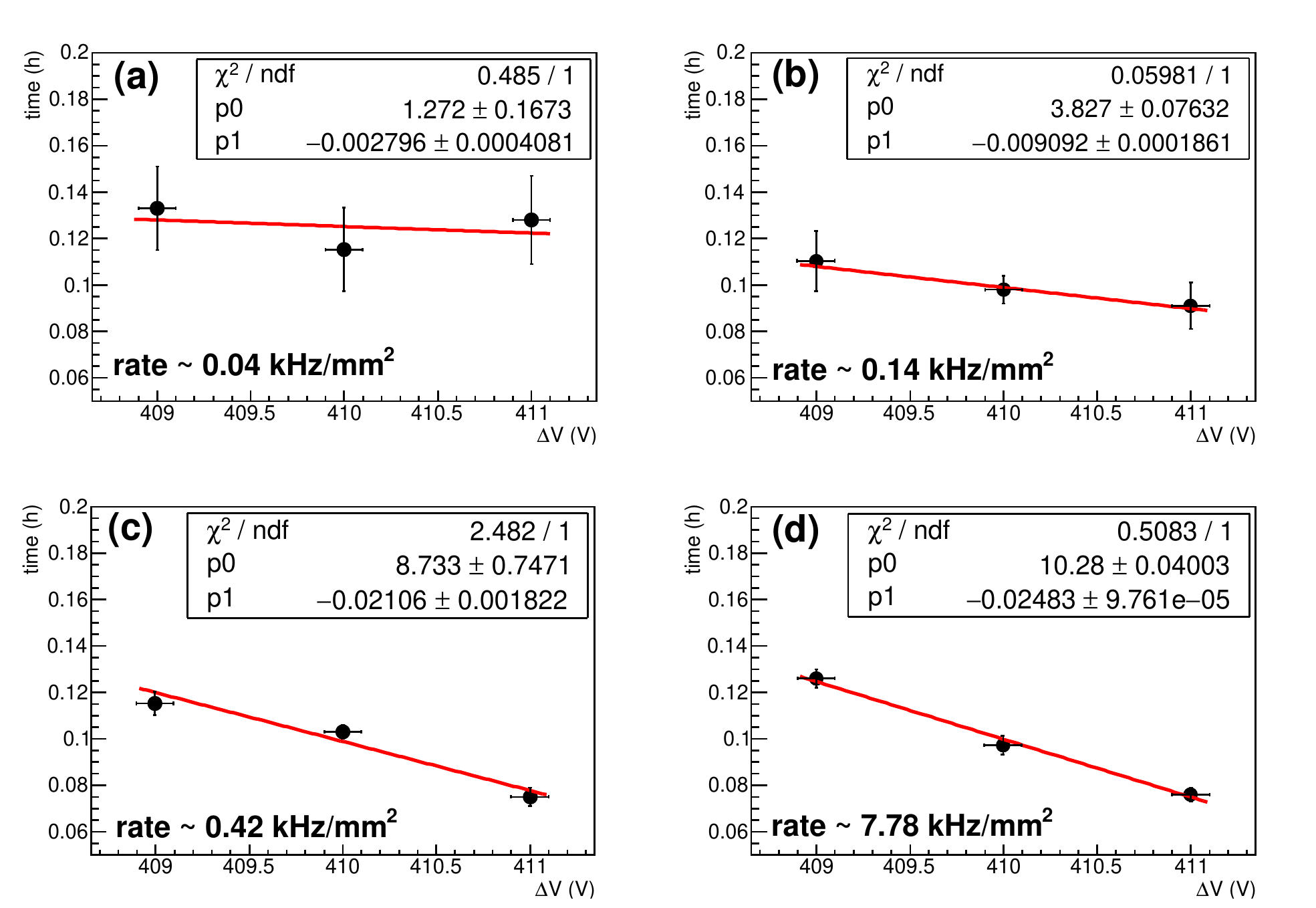}
			\caption{Variation of the initial gain decrease time as a function of voltage across the GEM foils~($\Delta V$) for different irradiation rates.} \label{fig5}
		\end{center}
	\end{figure}
\begin{figure}[htb!]
		\begin{center}
			\includegraphics[scale=0.405]{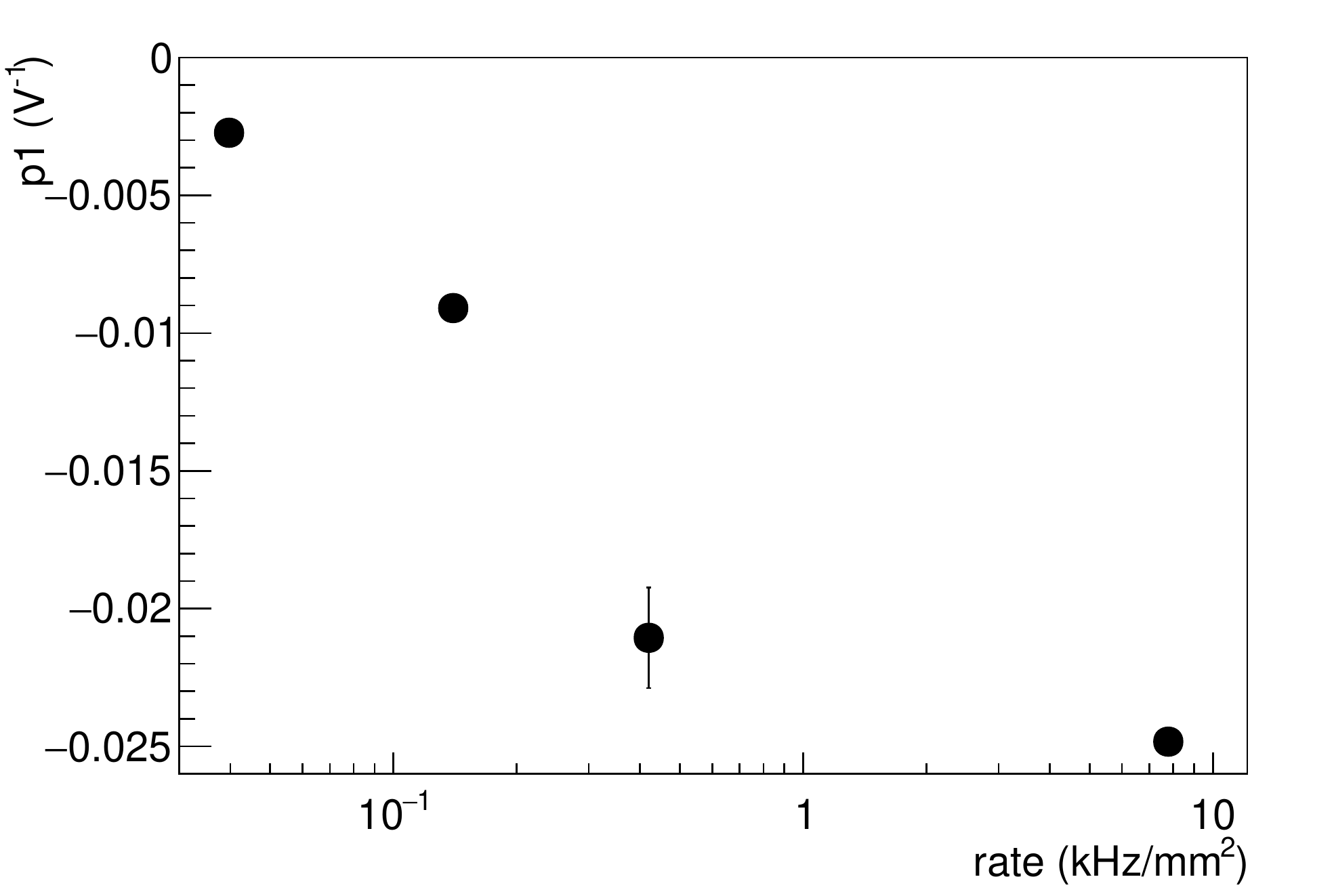}
			\caption{Variation of the slope~(p1) as a function of irradiation rates.} \label{slope_rate}
		\end{center}
	\end{figure}

	%%%%%%%%%%%%%%%%%%%%%%%%%%%%%%%%%%%%%%%%%%%%%%%%%%%
	In Fig.~\ref{fig5}, the variation in the time of the initial decrease of gain for different irradiation rates is shown. The data points are fitted with a linear function. It is observed that the time up to which the gain decreases initially due to the polarisation effect is anti-correlated with the voltage across the GEM foils. An effect of the irradiation rates on the polarisation effect is also observed. As shown in Fig.~\ref{slope_rate}, the rate of decrease of time with $\Delta$V increases with the increasing rate of irradiation.
	%%%%%%%%%%%%%%%%%%%%%%%%%%%%%%%%%%%%%%%%%%%%%%%%%%%
	%%%%%%%%%%%%%%%%%%%%%%%%%%%%%%%%%%%%%%%%%%%%%%%%%%%
	
	To observe the effect of charging-up phenomenon, the HV is switched on for around $\sim$ 60 minutes before starting the measurement to ensure that the polarisation of the dielectric in the GEM foil is over. After that, the measurement is started as soon as the Fe$^{55}$ X-ray source is placed on the chamber. The spectra for 60 seconds are stored at an interval of 120 seconds and analysed to obtain the gain of the chamber. Ambient temperature, pressure and humidity are also monitored continuously. To nullify the effect of temperature and pressure on the gain of the chamber, the gain is normalised with T/p and then fitted with the exponential function as discussed in Ref.~\cite{s_chatterjee_charging_up},  
	
	\begin{equation}\label{eqn1}
	G = p_0(1-p_1e^{-t/p_2})
	\end{equation}
	
	where \textit{G} is the normalised gain, \textit{p$_0$} and \textit{p$_1$} are the constants, \textit{t} is the measured time in hour and \textit{p$_2$} is the time constant of the charging-up effect. 
	%%%%%%%%%%%%%%%%%%%%%%%%%%%%%%%%%%%%%%%%%%%%%%%%%%%
	\begin{figure}[htb!]
		\begin{center}
			\includegraphics[scale=0.30]{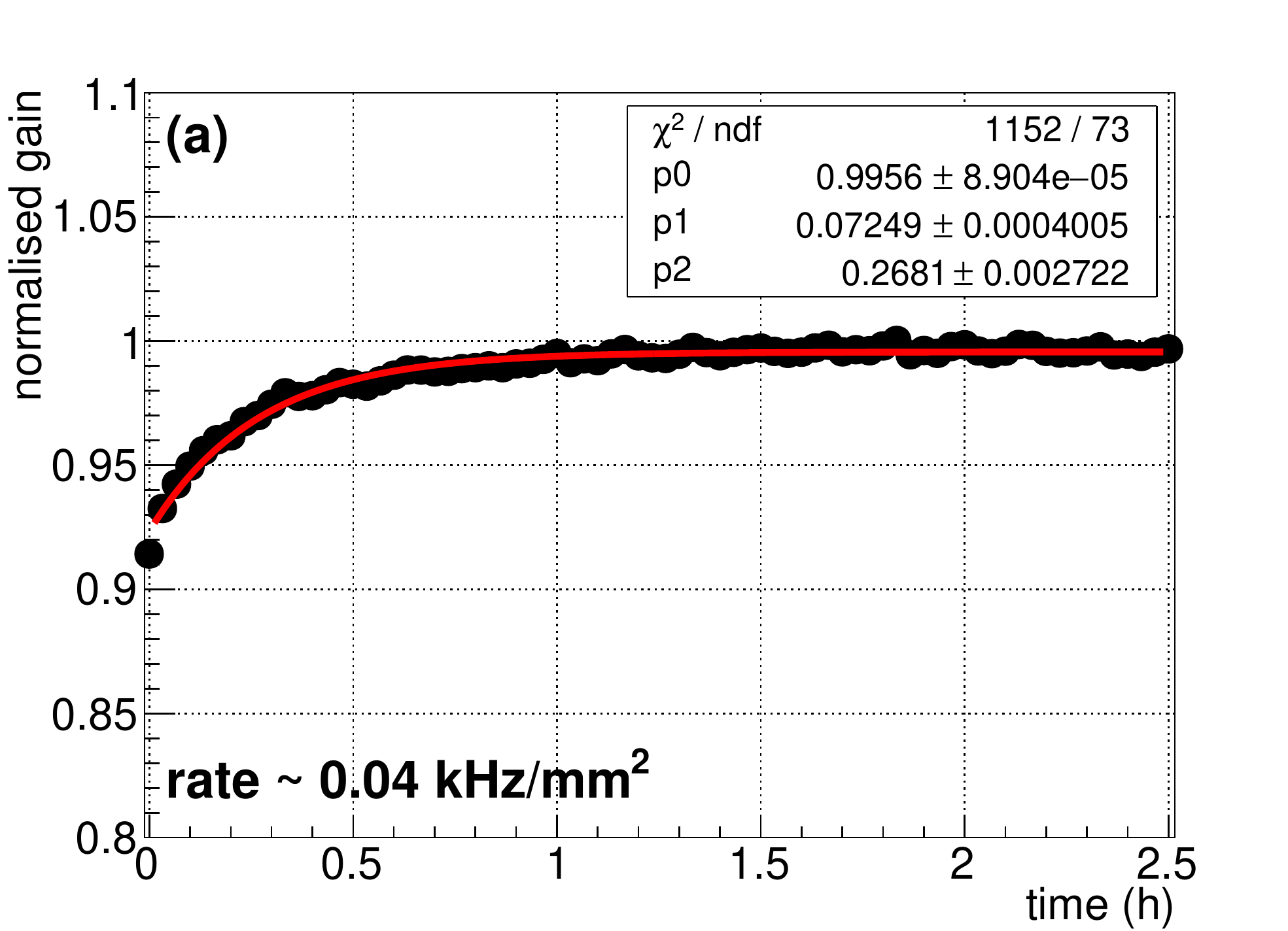}
			\includegraphics[scale=0.30]{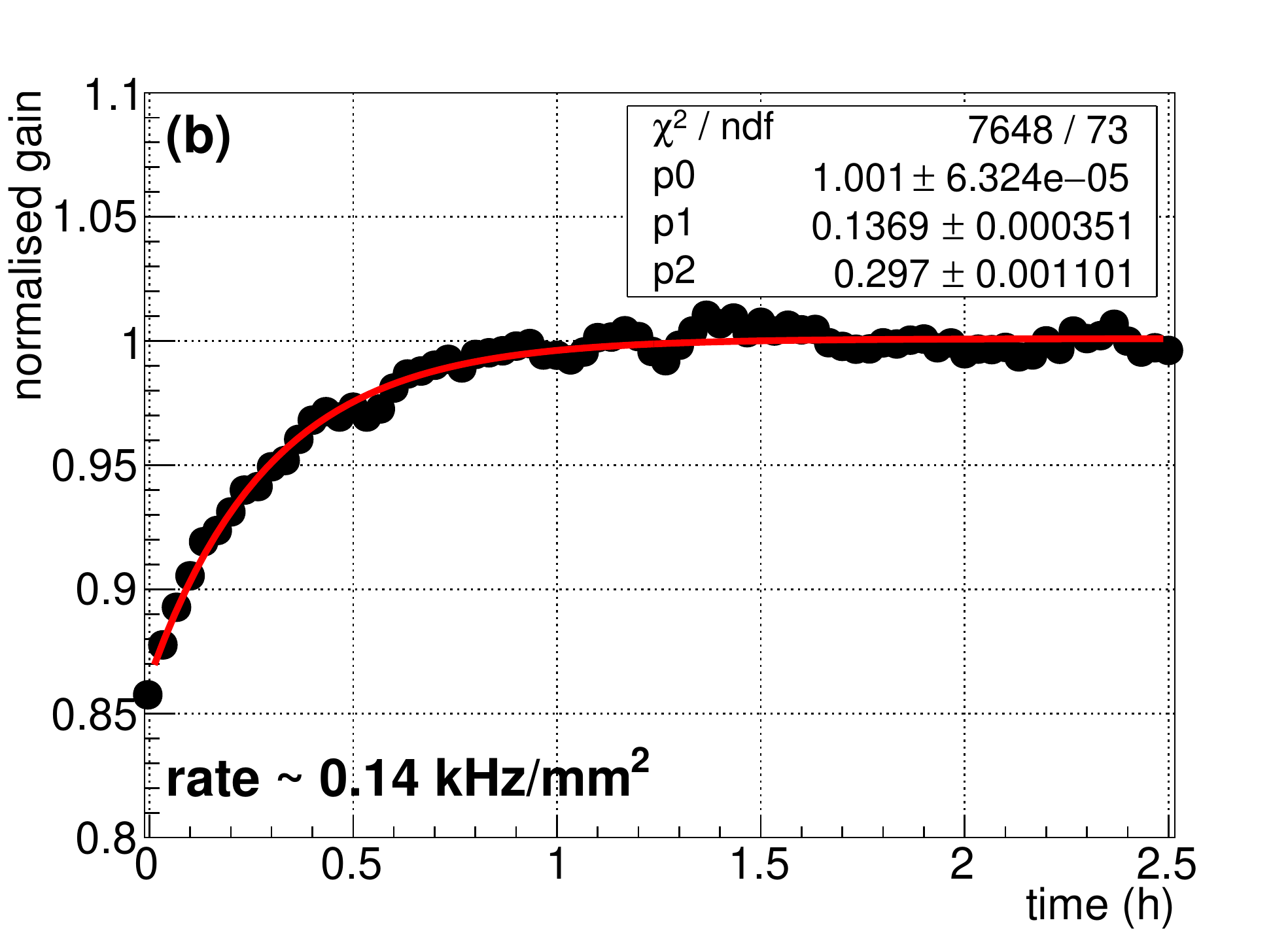}
			\includegraphics[scale=0.30]{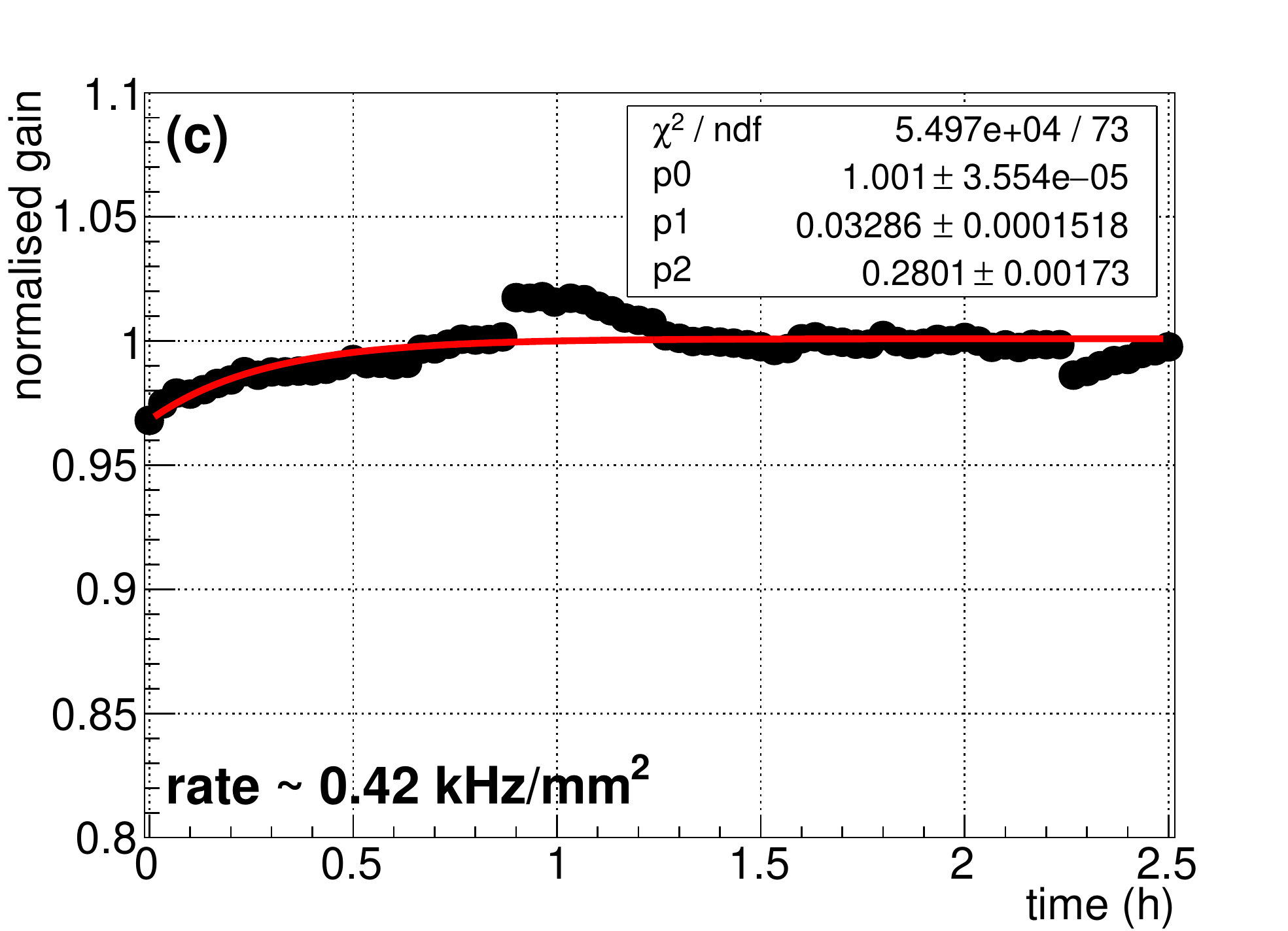}
			\includegraphics[scale=0.30]{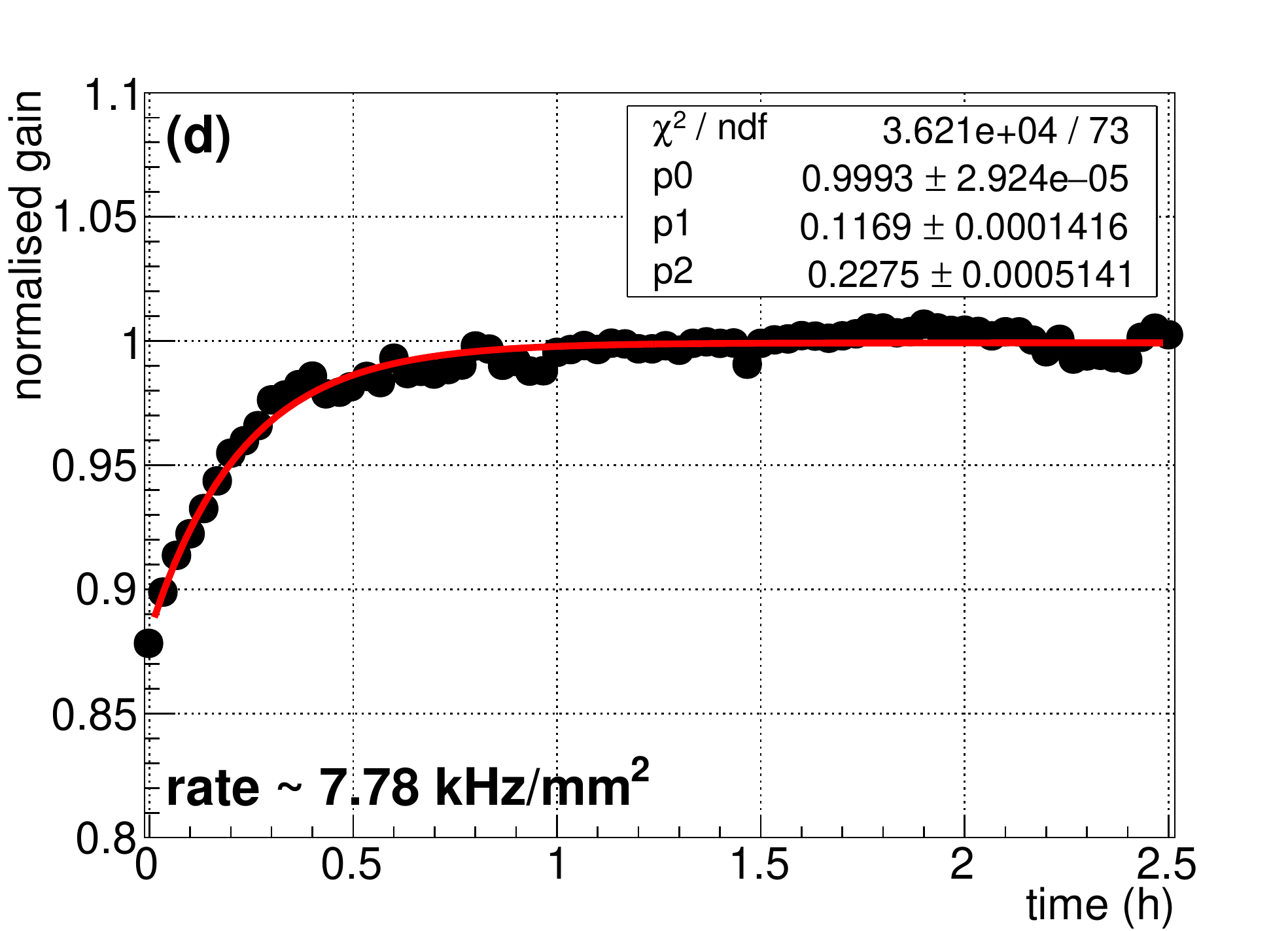}
			\caption{Variation of normalised gain as function of time (hour) for different irradiation rates. All the measurements are carried out at a HV of - 5085 V ($\Delta V = 409 V$).  }\label{fig6}
		\end{center}
	\end{figure}
	%%%%%%%%%%%%%%%%%%%%%%%%%%%%%%%%%%%%%%%%%%%%%%%%%%%
	In Fig.~\ref{fig6}, the variation of normalised gain as a function of time (hour) at a HV of - 5085 V is shown for different irradiation rates as listed in table~\ref{table3}. In the case of Fig.~\ref{fig6}(c), the small jumps in the normalised gain around 0.9~hour and 2.2~hour are due to some sudden change in the ambient T/p value as recorded by the data logger. 
	The details of the charging-up time, irradiation rates and saturated gain are listed in table~\ref{table3}. 
	%%%%%%%%%%%%%%%%%%%%%%%%%%%%%%%%%%%%%%%%%%%%%%%%%%%%%%%%%%%%%%%%%%%
	\begin{center}
		\begin{table}[htb!]
			\begin{center}
				\vspace*{0.4cm}
				\resizebox{\columnwidth}{!}{
				\begin{tabular}{|c|c|c|c|} \hline
					$\Delta V$ & rate & Saturated & Charging-up  \\ 
					(V) & (kHz/mm$^2$)& gain &time (h)\\ \hline
					& $\sim$ 0.04 &$\sim$ 13000  & 0.268 ($\underline{+}0.003$)  \\ 
					409& $\sim$ 0.14  &$\sim$ 11800  & 0.297 ($\underline{+}0.001$)   \\ 
					&  $\sim$  0.42&$\sim$ 12600 & 0.280 ($\underline{+}0.002$)  \\ 
					&  $\sim$  7.78 &$\sim$ 12200 & 0.228 ($\underline{+}0.001$)  \\ \hline
					&  $\sim$ 0.04 &$\sim$ 13800  & 0.285 ($\underline{+}0.005$)  \\ 
					410&  $\sim$ 0.14  &$\sim$ 13700  & 0.351 ($\underline{+}0.003$)   \\ 
					&  $\sim$  0.42&$\sim$ 13500 & 0.407 ($\underline{+}0.002$)  \\ 
					&  $\sim$  7.78 &$\sim$ 13400 & 0.335 ($\underline{+}0.002$)  \\ \hline
					
					&  $\sim$ 0.14  &$\sim$ 14500  & 0.190 ($\underline{+}0.002$)   \\ 
					411  & $\sim$  0.42&$\sim$ 14700 & 0.436 ($\underline{+}0.002$)  \\ 
					&  $\sim$  7.78 &$\sim$ 14000 & 0.186 ($\underline{+}0.001$)  \\ \hline
				\end{tabular}
				}
				\caption{Saturated gain and charging-up time for different irradiation rates.} \label{table3}
			\end{center}
			
		\end{table}
	\end{center}
	%%%%%%%%%%%%%%%%%%%%%%%%%%%%%%%%%%%%%%%%%%%%%%%%%%%%%%%%%%%%%%%%%%%
	%%%%%%%%%%%%%%%%%%%%%%%%%%%%%%%%%%%%%%%%%%%%%%%%%%%
	\begin{figure}[htb!]
		\begin{center}
			\includegraphics[scale=0.40]{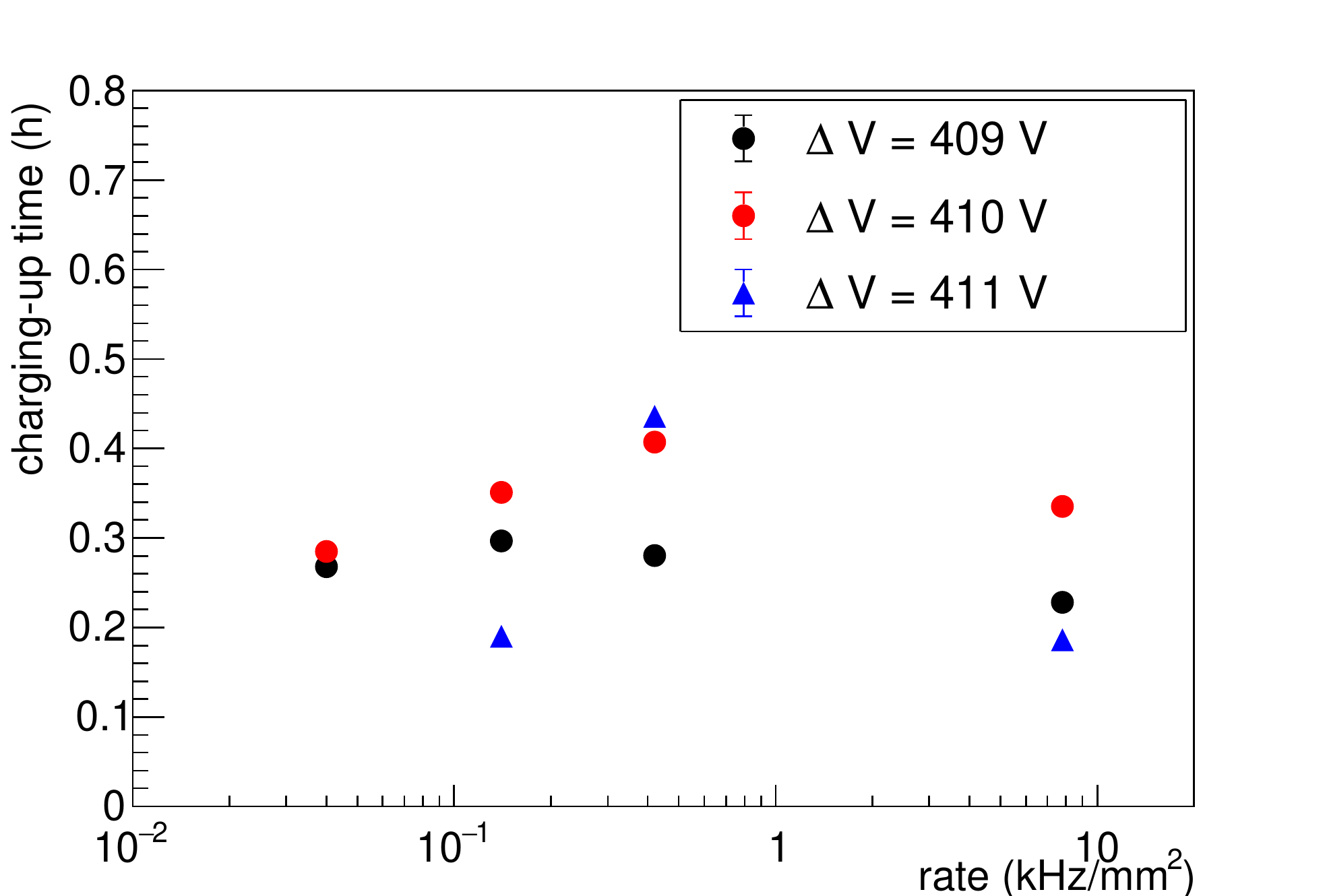}
			\caption{Charging-up time as a function of rate for different voltage settings. }\label{fig7}
		\end{center}
	\end{figure}
	%%%%%%%%%%%%%%%%%%%%%%%%%%%%%%%%%%%%%%%%%%%%%%%%%%% 
	
\begin{figure}[htb!]
		\begin{center}
			\includegraphics[scale=0.35]{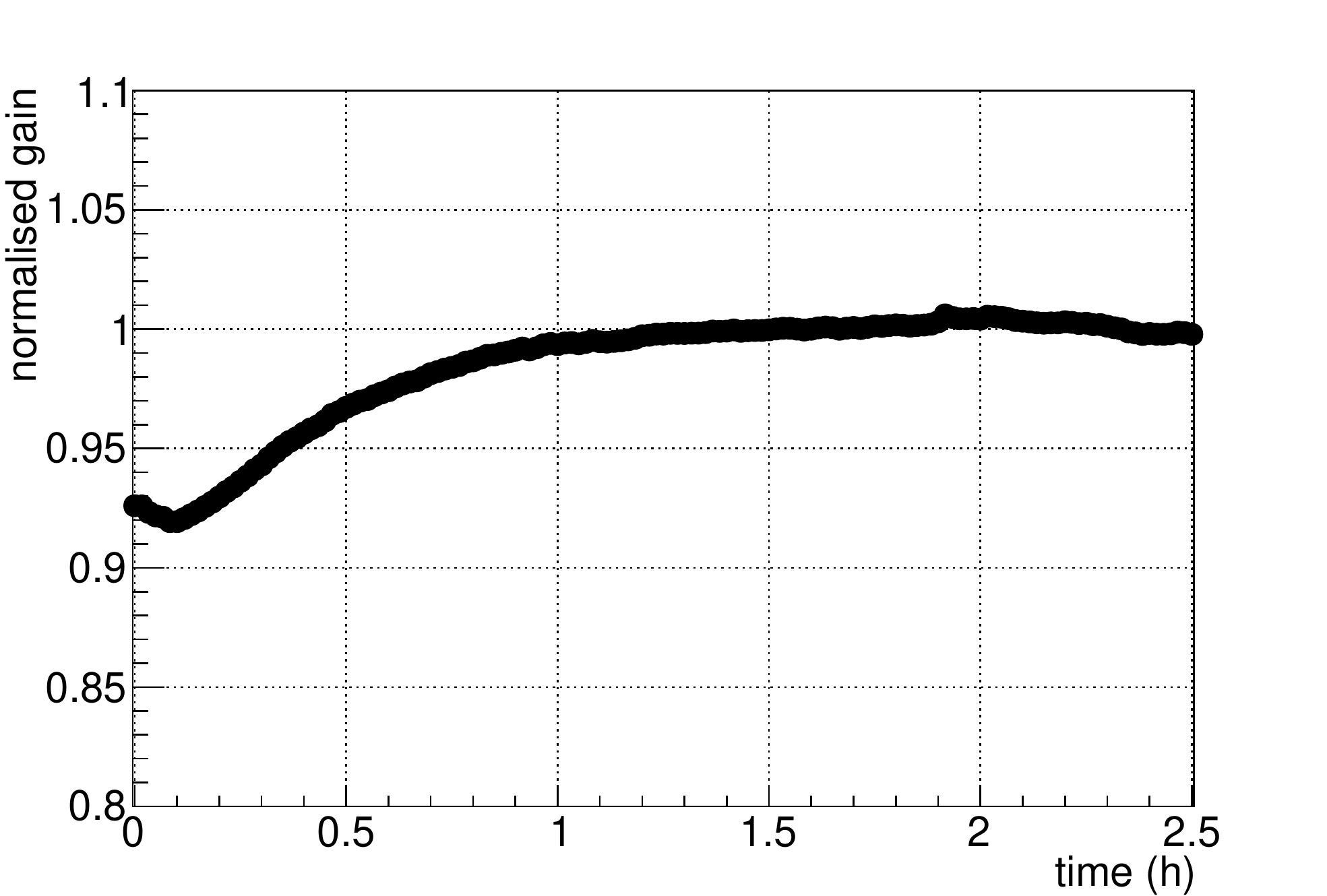}
			\caption{Variation of normalised gain as a function of time (in hour) at a HV of - 5085~V and particle flux $\sim$ 7.78 kHz/mm$^2$. The measurement is started as soon as the HV reached to its specific value and the source placed on the chamber. }\label{fig8}
		\end{center}
	\end{figure}

	The charging-up time for different X-ray flux is shown in Fig.~\ref{fig7}. The charging up time varies between 0.2~-~0.4 hour depending on the rate of irradiation as shown in Fig.~\ref{fig7}.
	In Fig.~\ref{fig8}, the variation of gain as a function of time~(hour) is shown to illustrate both the effects of polarisation and charging-up of the dielectric medium on the gain of the chamber. It is visible in Fig.~\ref{fig8} that initially, the gain decreases due to the initial polarisation effect of the dielectric for around 0.1 hour. After that due to the charging-up effect, the gain starts to increase and saturates after around 1.0 hour.
	%%%%%%%%%%%%%%%%%%%%%%%%%%%%%%%%%%%%%%%%%%%%%%%%%%%
	%%%%%%%%%%%%%%%%%%%%%%%%%%%%%%%%%%%%%%%%%%%%%%%%%%%  

	\section{Summary and Outlook}\label{summary}
	The effect of charging-up phenomena and initial polarisation of the dielectric inside the active volume of the 10~cm~$\times$~10~cm single mask triple GEM prototype is investigated with Fe$^{55}$ X-ray source at different gains and irradiation rates. The chamber is operated with Ar/CO$_2$ gas mixture in the 70/30 volume ratio. The effect of initial polarisation of the dielectric is investigated for four different irradiation rates and with three different voltage settings. To quantify the effect of rate and $\Delta V$ across the GEM foils on the polarisation effect, the initial decrease of the gain of the chamber is fitted with a 2$^{nd}$ degree polynomial and then from the fitting parameters, the time up to which the gain is decreasing initially is found out. It is observed that this time decreases with increasing $\Delta V$ across the GEM foils. A correlation is also observed between the irradiation rates and the effect of the initial polarisation of the dielectric. At a given $\Delta$V, with increasing particle flux, the time required to reach the minimum gain value reduces.  After the initial polarisation effect, the gain of the chamber increases due to the modification of the electric field lines inside the GEM holes, \textit{i.e} the charging-up effect. 
	In order to quantify the effect of the rate on the charging-up effect, the T/p normalised gain is fitted with eqn.~\ref{eqn1} and $\textit{p2}$ gives the charging-up time taking analogy from the charging-up mechanism in RC networks~\cite{RC_network}. The charging-up time is found to be between 0.2-0.4 hour. 
	
	The comparison between the charging-up effect measurement between the double mask triple GEM chamber as reported earlier~\cite{s_chatterjee_charging_up} and the present measurement with the single mask triple GEM chamber are the following. The HV filter was not used before the resistive chain in the measurement setup with a double mask triple GEM chamber. The charging-up measurement was performed at a $\Delta V$ $\sim$ 390 V across GEM foils and the respective drift field, transfer fields, and induction field was $\sim$~2.3 kV/cm, $\sim$~3.5 kV/cm, and $\sim$~3.5 kV/cm. The charging-up time was found to be 2.376~$\underline{+}$~0.02~hour, 1.524~$\underline{+}$~0.008~hour and 1.395~$\underline{+}$~0.004 hour for irradiation rates of $\sim$~0.08 kHz/mm$^2$, 0.2 kHz/mm$^2$ and 3.2 kHz/mm$^2$ respectively. The saturation gain for the irradiation rates $\sim$~0.08 kHz/mm$^2$, 0.2 kHz/mm$^2$ and 3.2~kHz/mm$^2$ was $\sim$ 4900, 5100 and 5500 respectively.

	\section{Acknowledgements}
	The authors would like to thank the RD51 collaboration for the support in building and initial	testing of the chamber in the RD51 laboratory at CERN. We would like to thank Dr.~L.~Ropelewski, Dr.~E. Oliveri of CERN and Prof.~Sibaji Raha, Prof.~Rajarshi Ray and Dr.~Sidharth K. Prasad of Bose Institute for valuable discussions and suggestions in the course of the study. This work is partially supported by the research grant SR/MF/PS-01/2014-BI from DST, Govt. of India, and the research grant of CBM-MuCh project from BI-IFCC, DST, Govt. of India. S. Biswas acknowledges the support of the DST-SERB Ramanujan Fellowship (D.O.No. SR/S2/RJN-02/2012).


\noindent
	\begin{thebibliography}{50}
		\bibitem{sauli_GEM}
		F. Sauli, Nucl. Instrum. Methods Phys. Res. A 386 (1997) 531.
		
		\bibitem{gem_review}
		A.F. Buzulutskov, Instrum Exp Tech 50, (2007) 287.
		
		\bibitem{ketzer}
		B. Ketzer et al., Nucl. Instrum. Meth. A 535 (2004) 314.
		
		\bibitem{CMS_upgrade}
		D. Abbaneo et al., Nucl. Instrum. Meth. A 718 (2013) 383.
		
		\bibitem{alice_upgrade}
		B. Ketzer, Nucl. Instrum. Meth. A 732 (2013) 237.
		
		
		\bibitem{cbm_detector_system}
		T. Balog, J. Phys. Conf. Ser. 503 (2014) 012019.
		
		\bibitem{s_biswas_spark}
		S. Biswas et al., Nucl. Instrum. Meth. A 800 (2015) 93.
		
		\bibitem{s_chatterjee_spark}
		S. Chatterjee et al., Nucl. Instrum. Meth. A 977 (2020) 164334.
		
		\bibitem{GEM_foil}
		Oliveira et al., United States Patent, Patent No.: US 8,597,490 B2.
		
		\bibitem{pinto_large_area_gem}
		S. Duarte Pinto et al., J. Instrum. 4 (2009) P12009.
		
		\bibitem{gem_hole_geometry_asym_1}
		A. Karadzhinova et al., J. Instrum. 10 (2015) P12014.
		
		\bibitem{s_das}
		S. Das, Nucl. Instrum. Methods Phys. Res. A 824 (2016) 518.
		
		\bibitem{gem_hole_geometry_asym_2}
		A. Shah et al., Nucl. Instrum. Methods Phys. Res. A 936 (2019) 459. 
		
		\bibitem{charging_up_philip}
		P. Hauer et al., Nucl. Instrum. Methods Phys. Res. A 976 (2020) 164205.
		
		\bibitem{charging_up_azmoun}
		B. Azmoun et al., IEEE Nuclear Science Symposium Conference Record VOL. 6 (2006) 3847.
		
		\bibitem{charging_up_alfonsi}
		M. Alfonsi, Nucl. Instrum. Methods Phys. Res. A	671 (2012) 6.
		
		\bibitem{s_chatterjee_charging_up}
		S. Chatterjee et al., J. Instrum. 15 (2020) T09011.
		
		
		\bibitem{RD51}
		http://rd51-public.web.cern.ch/rd51-public/.
		
		\bibitem{rama_adak}
		R. P. Adak et al., J. Instrum. 11 (2016) T10001.
		
		
		\bibitem{preamplifier}
		CDT CASCADE Detector Technologies GmbH, Germany,\\ 
		www.n-cdt.com.
		
		\bibitem{data_logger}
		S. Sahu et al., J. Instrum. 12 (2017) C05006.
		
		\bibitem{tp_gem}
		M. C. Altunbas et al., Nucl. Instrum. Methods Phys. Res. A 515 (2003) 249.
		
		\bibitem{cern_root}
		https://root.cern/.
		
		\bibitem{RC_network}
		V. Tikhonov et al., Nucl. Instrum. Methods Phys. Res. A 478 (2002) 452
		
	\end{thebibliography}
\end{document}